\documentclass{JAC2001}  

\usepackage{graphicx}
\newcommand{\ud}{\mathrm{d}}
\begin{document}

\title{NONLINEAR DYNAMICS OF HIGH-BRIGHTNESS BEAMS}
\author{Antonina N.  Fedorova, Michael G. Zeitlin\\ 
IPME, RAS, V.O. Bolshoj pr., 61, 199178, St.~Petersburg, Russia
\thanks{e-mail: zeitlin@math.ipme.ru}\thanks{ http://www.ipme.ru/zeitlin.html;
http://www.ipme.nw.ru/zeitlin.html}   
}
\maketitle

\begin{abstract}
The consideration of transverse dynamics of relativistic spa\-ce-char\-ge             
dominated beams and halo growth due to bunch oscillations 
is based on variational approach to rational (in dynamical     
variables) approximation for rms envelope equations. It allows to           
control contribution from each scale of underlying multiscales                 
and represent solutions via exact nonlinear eigenmodes expansions.        
Our approach is                                                                    
based on methods provided possibility to work with well-localized bases        
in phase space and good convergence properties of the corresponding                
expansions.                                                      
\end{abstract}

\section{Introduction}
In this paper we consider the applications of a new nu\-me\-ri\-cal\--analytical 
technique based on the methods of local nonlinear harmonic
analysis or wavelet analysis to nonlinear rms envelope dynamics
problems which can be characterized by  collective type behaviour [1], [2].
Such approach may be useful in all models in which  it is 
possible and reasonable to reduce all complicated problems related with 
statistical distributions to the problems described 
by systems of nonlinear ordinary/partial differential 
equations with or without some (functional) constraints.
Wavelet analysis is a set of mathematical
methods, which gives the possibility to work with well-localized bases in
functional spaces and gives the maximum sparse forms for the general 
type of operators (differential, integral, pseudodifferential) in such bases. 
Our approach is based on the 
variational-wavelet approach from [3]-[14],
that allows to consider rational type of 
nonlinearities in rms dynamical equations.
The solution has the multiscale/multiresolution decomposition via 
nonlinear high-localized eigenmodes,
which corresponds to the full multiresolution expansion in all underlying time/space 
scales. 
We may move
from coarse scales of resolution to the 
finest one for obtaining more detailed information about our dynamical process.
In this way we give contribution to our full solution
from each scale of resolution or each time/space scale or from each nonlinear eigenmode. 
The same is correct for the contribution to power spectral density
(energy spectrum): we can take into account contributions from each
level/scale of resolution.
Starting  in part 2 from general rms envelope dynamics
model
we consider in part 3 the approach based on
variational-wavelet formulation. 
We give explicit representation for all dynamical variables in the base of
compactly supported wavelets or nonlinear eigenmodes.  Our solutions
are parametrized
by the solutions of a number of reduced algebraical problems one from which
is nonlinear with the same degree of nonlinearity and the others  are
the linear problems depend on particular wavelet type approach.
In part 4 we consider numerical modelling based on our analytical approach.

\section{RMS EQUATIONS}

We  consider an approach based on 
the second moments of the distribution functions for  the calculation
of evolution of rms envelope of a beam.
The rms envelope equations are the most useful for analysis of the 
beam self--forces (space--charge) effects and also 
allow to consider  both transverse and longitudinal dynamics of
space-charge-dominated relativistic high--brightness
axisymmetric/asymmetric beams, which under short laser pulse--driven
radio-frequency photoinjectors have fast transition from nonrelativistic
to relativistic regime [1]. The analysis of halo growth in beams, appeared
as result of bunch oscillations in the particle-core model, is also based
on three-dimensional envelope equations [1], [2]. 
Let $f(x_i)$ be the distribution function, which gives full information
about 
noninteracting ensemble of beam particles regarding to trace space or 
transverse phase coordinates $(x_i,x_j)$. 
Then we may extract the first nontrivial effects of collective dynamics from 
the second moments
\begin{eqnarray}
\sigma_{x_i x_j}^2=<x_i x_j>=\int\int x_i x_j f(x)\ud x_i\ud x_j 
\end{eqnarray}
RMS emittances are given by
\begin{equation}
\varepsilon^2_{x_i,rms}=<x_i^2><\dot{x}_i^2>-<x_i \dot{x}_i>^2 
\end{equation}
We consider the following most general
case of rms envelope equations, which describe evolution
of the moments
in the model of halo formation
by bunch oscillations (ref. [2] for full designation):
\begin{eqnarray}
\ddot\sigma_x+k_x^2(s)\sigma_x-\frac{\xi_x}{\sigma_y\sigma_z}-\frac{\varepsilon^2_x}{\sigma_x^3}&=&0,\nonumber\\
\ddot\sigma_y+k_y^2(s)\sigma_y-\frac{\xi_y}{\sigma_x\sigma_z}-\frac{\varepsilon^2_y}{\sigma_y^3}&=&0,\\
\ddot\sigma_z+k_z^2(s)\sigma_z-\gamma^2\frac{\xi_z}{\sigma_x\sigma_y}-\frac{\varepsilon^2_z}{\sigma_z^3}&=&0,\nonumber
\end{eqnarray}
where $\sigma_x(s), \sigma_y(s), \sigma_z(s)$ are bunch envelopes, $\xi_x, \xi_y$, 
$\xi_z= F(\sigma_x,\sigma_y, \sigma_z)$.
After transformations to Cauchy form we can see that
all these equations from the formal point of view are not more than
ordinary differential equations with rational nonlinearities
and variable coefficients.
Also, we consider regimes in which we are interested in constraints on emittances:
\begin{equation}
\varepsilon^2_{x_i,rms}=c_i,
\end{equation}
where $c_i$ are constants.
In the same way according to [2] we may consider the case of energy-type functional-differential
constraints on  emittances.
A different approach is considered in our related paper in this Proceedings [15].

\section{RATIONAL DYNAMICS WITH CONSTRAINTS}
                                                           
Our problems above may be formulated as the systems of ordinary differential            
equations                                                               
\begin{eqnarray}\label{eq:pol0}                                
& & Q_i(x)\frac{\ud x_i}{\ud t}=P_i(x,t),\quad x=(x_1,..., x_n),\\
& &i=1,...,n, \quad                                                                        
 \max_i  deg \ P_i=p, \quad \max_i deg \  Q_i=q \nonumber                  
\end{eqnarray}                                                 
with initial (or boundary) conditions $x_i(0)$, $x_i(T)$ and  $P_i, Q_i$ are not more    
than polynomial functions of dynamical variables $x_j$                                 
and  have arbitrary dependence on time/length parameter. 
Of course, we consider such $Q_i(x)$ which do not lead to the singular
problem with $Q_i(x)$, when $t=0$ or $t=T$, i.e. $Q_i(x(0)), Q_i(x(T))\neq\infty$, 0.
We'll consider these  equations as the following operator equation.
Let $L$ be an arbitrary nonlinear (rational) matrix differential operator of the first order with matrix dimension d
(d=6 in our case) corresponding to the system of equations (5), 
which acts on some set of functions
$\Psi\equiv\Psi(t)=\Big(\Psi^1(t),\dots,\Psi^d(t)\Big), \quad t \in\Omega\subset R$
from $L^2(\Omega)$:
\begin{equation}
L\Psi\equiv L(R,t)\Psi(t)=0,
\end{equation}
where
$R\equiv R(t,\partial /\partial t, \Psi)$.

Let us consider now the N mode approximation for solution as the following ansatz (in the same way
we may consider different ansatzes):
\begin{equation}
\Psi^N(t)=\sum^N_{r=1}a^N_{r}\psi_r(t)
\end{equation}
We shall determine the coefficients of expansion from the following variational conditions
(different related variational approaches are considered in [3]-[14]):
\begin{equation}
L^N_{k}\equiv\int(L\Psi^N)\psi_k(t)\ud t=0
\end{equation}
We have exactly $dN$ algebraical equations for  $dN$ unknowns $a_{r}$.
So, variational approach reduced the initial problem (5) or (6) to the problem of solution 
of functional equations at the first stage and some algebraical problems at the second
stage. 
Here $\psi_k(t)$ are useful basis functions of  some functional
space ($L^2, L^p$, Sobolev, etc) corresponding to concrete
problem. 
As result we have the following reduced algebraical system
of equations (RSAE) on the set of unknown coefficients $a_i^N$ of
expansions (7):
\begin{eqnarray}\label{eq:pol2}
L(Q_{ij},a_i^N,\alpha_I)=M(P_{ij},a_i^N,\beta_J),
\end{eqnarray}
where operators L and M are algebraization of RHS and LHS of initial problem
(\ref{eq:pol0}).
$Q_{ij}$ are the coefficients (with possible time dependence) of LHS of initial
system of differential equations (\ref{eq:pol0}) and as consequence are coefficients
of RSAE.
 $P_{ij}$ are the coefficients (with possible time dependence) of RHS
of initial system of differential equations (\ref{eq:pol0}) and as consequence
are the coefficients of RSAE.
$I=(i_1,...,i_{q+2})$, $ J=(j_1,...,j_{p+1})$ are multiindexes, by which are
labelled $\alpha_I$ and $\beta_I$,  the other coefficients of RSAE (\ref{eq:pol2}):
\begin{equation}\label{eq:beta}
\beta_J=\{\beta_{j_1...j_{p+1}}\}=\int\prod_{1\leq j_k\leq p+1}\psi_{j_k},
\end{equation}
where p is the degree of polynomial operator P (\ref{eq:pol0})
\begin{equation}\label{eq:alpha}
\alpha_I=\{\alpha_{i_1}...\alpha_{i_{q+2}}\}=\sum_{i_1,...,i_{q+2}}\int
\psi_{i_1}...\dot{\psi_{i_s}}...\psi_{i_{q+2}},
\end{equation}
where q is the degree of polynomial operator Q (\ref{eq:pol0}),
$i_\ell=(1,...,q+2)$, $\dot{\psi_{i_s}}=\ud\psi_{i_s}/\ud t$.

According to [3]-[14] we may extend our approach to the case when we have additional
constraints (4) on the set of our dynamical variables $\Psi$ or $x$.
In this case by using the method of Lagrangian multipliers we again may apply the same approach but
for the extended set of variables. As result we receive the expanded system of algebraical equations
analogous to the system (9). Then, after reduction we again can extract from its solution the coefficients 
of expansion (7).  
Now, when we solve RSAE (\ref{eq:pol2}) and determine
unknown coefficients from formal expansion (7) we therefore
obtain the solution of our initial problem.
It should be noted if we consider only truncated expansion (7) with N terms
then we have from (\ref{eq:pol2}) the system of $N\times d$ algebraical equations
with degree $\ell=max\{p,q\}$
and the degree of this algebraical system coincides
with degree of initial differential system.
So, we have the solution of the initial nonlinear
(rational) problem  in the form
\begin{eqnarray}\label{eq:pol3}
x(t)=x(0)+\sum_{k=1}^Na_k^N \psi_k(t),
\end{eqnarray}
where coefficients $a_k^N$ are roots of the corresponding
reduced algebraical (polynomial) problem RSAE (\ref{eq:pol2}).
Consequently, we have a parametrization of solution of initial problem
by solution of reduced algebraical problem (\ref{eq:pol2}).

The problem of
computations of coefficients $\alpha_I$ (\ref{eq:alpha}), $\beta_J$
(\ref{eq:beta}) of reduced algebraical
system
may be explicitly solved in wavelet approach.
The obtained solutions are given
in the form (\ref{eq:pol3}),
where
$\psi_k(t)$ are wavelet basis functions. In our case $\psi_k(t)$
are obtained via multiresolution expansions and represented by
compactly supported wavelets. 
Because affine
group of translation and dilations is inside the approach, this
method resembles the action of a microscope. We have contribution to
final result from each scale of resolution from the whole
infinite scale of spaces:
$$
...V_{-2}\subset V_{-1}\subset V_0\subset V_{1}\subset V_{2}\subset ...,
$$
where the closed subspace
$V_j (j\in {\bf Z})$ corresponds to  level j of resolution, or to scale j.
This multiresolution functional space decomposition corresponds to exact nonlinear
eigenmode decompositions (12).

It should be noted that such representations 
give the best possible localization
properties in the corresponding (phase)space/time coordinates. 
In contrast with different approaches formulae (7), (12) do not use perturbation
technique or linearization procedures 
and represent dynamics via generalized nonlinear localized eigenmodes expansion.  
So, by using wavelet bases with their good (phase)space/time      
localization properties we can construct high-localized (coherent)  structures in      
spa\-ti\-al\-ly\--ex\-te\-nd\-ed stochastic systems with collective behaviour.
      
\begin{figure}[htb]                                                             
\centering                                                                       
\includegraphics*[width=60mm]{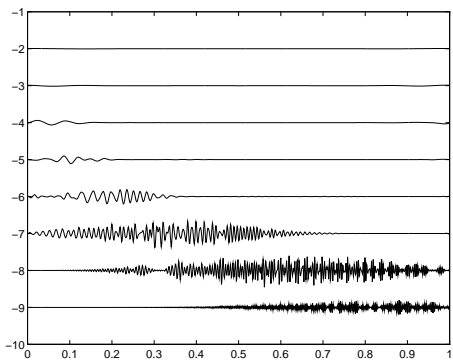}
\caption{Multiscale decomposition.}                              
\end{figure}                                                                   
                                                                                
\begin{figure}[htb]                                                            
\centering                                                                      
\includegraphics*[width=60mm]{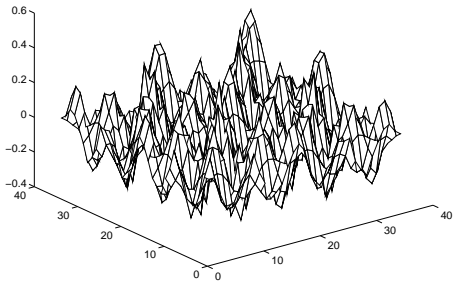}
\caption{$\sigma_x$--$\sigma_y$ section.}                                           
\end{figure}                                                                  

\section{Modelling}

So, our N mode construction (7), (12) gives the following representation for solution of rms equations (3):
\begin{eqnarray}
x(t)&=&x_{N}^{slow}(t)+\sum_{i\ge N}x^i(\omega_it),
\quad \omega_i \sim 2^i
\end{eqnarray}
where $x^r(s)$ may be represented by some family of (nonlinear)
eigenmodes and gives as a result the multiresolution/multiscale representation in the
high-localized wavelet bases.
The corresponding decomposition is  presented on Fig.~1 and two-dimensional transverse 
section $\sigma_x$--$\sigma_y$ on  Fig.~2. 

\section{Acknowledgments}

We would like to thank The U.S. Civilian Research \& Development Foundation (CRDF) for
support (Grants TGP-454, 455), which gave us the possibility to present our nine papers during
PAC2001 Conference in Chicago and Ms.Camille de Walder from CRDF for her help and encouragement.

 \end{document}